\documentclass[twoside]{dis09}
\usepackage[latin1]{inputenc}
\usepackage[dvips]{graphicx,epsfig,color}
\usepackage{wrapfig,rotating}
\usepackage{amssymb,amsmath,array}

\newcommand{\beeq}{\begin{eqnarray}}
\newcommand{\eeeq}{\end{eqnarray}}
\newcommand{\be}{\begin{equation}}
\newcommand{\ee}{\end{equation}}
\newcommand{\bea}{\begin{array}}
\newcommand{\eea}{\end{array}}

\newcommand{\eq}{&=&}
\newcommand{\eto}{{\mbox{\textrm e}}}

\def\xp{x_{{I\!\!P}}}

\def\qbar{\overline{q}}
\def\cbar{\overline{c}}
\def\bbar{\overline{b}}
\def\sigmahat{\hat{\sigma}}

\def\gev{{\textrm GeV}}

\begin{document}
\title{Precise dipole model analysis of diffractive DIS}

\author{Agnieszka \L{}uszczak$^1$ 
%
\thanks{This work is partially supported by the grants of MNiSW Nos. N202 246635 and
N202 249235.}
%
\vspace{.3cm}\\
%
1- Institute of Nuclear Physics Polish Academy of Sciences\\
Radzikowskiego 152, Cracow, Poland}
%


\maketitle

\begin{abstract}
We analyse the newest diffractive deep inelastic scattering data from HERA using
the dipole model approach. We find a reasonable good agreement between the predictions and the data although
the region of small values of a kinematic variable $\beta$
needs refinement. A way to do this is to consider an approach with diffractive
parton distributions evolved with the DGLAP evolution equations.
\end{abstract}

\section{Introduction}
The most promissing QCD based approach to DIS diffraction is formulated by a systematic formation of the diffractive state from  parton components of the light cone virtual photon wave function, projected onto the color singlet state. The lowest order  states are formed by a quark-antiquark pair $(q\qbar))$ and a $q\qbar$--gluon system while higher order states
contain aditional $q\qbar$ pairs and gluons $g$. We will concentrate on the first two components since they can be viewed 
in the configuration space conjugate to parton transverse momentum
as quark or gluon color dipoles. This is the basis of the dipole models  which have to be supplemented by the way the dipoles interact with the proton, which is described by the
scattering amplitude $N(r,b)$.
In this analysis, we consider two  important parameterisations of the dipole scattering amplitude, 
called GBW  \cite{Golec-Biernat:1998js} and CGC  \cite{Soyez:2007kg}, in which parton saturation results are built in.
Here we present a  precise comparison of the results of the dipole models
with these two parameterisations with  the newest data from HERA  on  the combination of the diffractive structure functions, obtained by the ZEUS \cite{Chekanov:2008cw} collaboration.

\subsection{Diffractive structure functions in dipole models}
In the dipole approach to DDIS, the diffractive structure function
$F_2^{D}$ is a sum of components corresponding to different diffractive final states produced by 
transversely $(T)$ and longitudinally $(L)$ polarised virtual photon  \cite{Bartels:1998ea}. 
We consider two component diffractive final state which is built from $q\qbar$ pair from transverse and longitudinal photon and $q\qbar g$ system from transverse photon.
Thus,  the structure function is given as a sum
\be\label{eq:1}
F_2^{D}(\xp,\beta,Q^2)=F_T^{(q\qbar)}+F_L^{(q\qbar)}+F_T^{(q\qbar g)}
\ee
where the kinematic variables depend on diffractive mass $M$ and center-of-mass energy of the $\gamma^* p$ system
$W$ through
\be\label{eq:2}
\xp = \frac{M^2+Q^2}{W^2+Q^2}\,,~~~~~~~~~~\beta=\frac{Q^2}{Q^2+M^2}
\ee
while the standard  Bjorken variable $x=\xp\beta$. The dependence of $F_2^D$ on the momentum transfer $t=(p-p^\prime)^2$ is integrated out.
The $q\qbar$ components from  transversely and longitudinally polarised  photons are given by 
\beeq\nonumber
\label{eq:5a}
\xp F_T^{(q\qbar)}\!\!\eq\!\!
\frac{3 Q^4}{64\/\pi^4\beta B_d}\,\sum_f e_f^2\int_{z_{f}}^{1/2}
dz\, z(1-z)
\\
&\times&\left\{
[z^2+(1-z)^2]\,Q^2_f\,\phi_1^2 \,+\,m_f^2\, \phi_0^2
\right\}~~~~~~~~
\\\nonumber
\\\label{eq:5}
\xp F_L^{(q\qbar)}\!\!\eq\!\!
\frac{3 Q^6}{16\/\pi^4\beta B_d}\,\sum_f e_f^2\int_{z_{f}}^{1/2}
dz\,z^3(1-z)^3\,\phi_0^2
\eeeq
where $f$ denotes quark flavours, $m_f$ is quark mass and the diffractive slope $B_d$ in the denominator results from
the $t$-integration of  the structure functions, assuming an exponential form for this dependence.
From HERA data, $B_d=6~\gev^{-2}$.
The variables
\be 
z_{f} = {\textstyle{\frac{1}{2}}}(1-\sqrt{1-4m_f^2/M^2})\,,~~~~
Q^2_f=z(1-z)Q^2+m_f^2
\ee 
and the functions  $\phi_i$  take the following form for $i=0,1$
\be\label{eq:8}
\phi_i
=\int_0^\infty dr r K_i\!\left(Q_f r\right)
J_i\!\left(k_fr\right) \sigmahat(\xp,r)
\ee
where $k_f=\sqrt{z(1-z)M^2-m_f^2}$ is the quark transverse momentum while
$K_i$ and $J_i$ are the Bessel functions.
The lower integration limit $z_f$ in eqs.~(\ref{eq:5a}) and (\ref{eq:5}) corresponds to a minimal value
of $z$ for which the diffractive  state with mass $M$ can be produced. In such a case $k_f=0$. 
At the threshold for the massive quark production  $M^2=4m_f^2$ and $z_f=1/2$, leading to $F_{T,L}^{(q\qbar)}=0$.
For massless quarks $z_f=0$.
The $q\qbar g$ diffractive component from transverse photons, computed for massless quarks is given by
\beeq\nonumber
\xp F_T^{(q\qbar g)}
\eq
\frac{81 \beta\alpha_s }{512\pi^5 B_d}\;
\sum_f e_f^2 
\int_\beta^1 \frac{dz}{(1-z)^3} 
\\\nonumber
&\times&\left[\left(1-\frac{\beta}{z}\right)^2+\left(\frac{\beta}{z}\right)^2\right]
\\\nonumber
\\\label{eq:10}
&\times&
\int_0^{(1-z)Q^2} dk^2 \log\left(\frac{(1-z)Q^2}{k^2}\right)\phi_2^2
\eeeq
where the function $\phi_{2}$ takes to form
\be\label{eq:11}
\phi_{2}
=k^2
\int_0^\infty dr\, r\, K_{2}\!\left(\sqrt{\frac{z}{1-z}}kr\right)\,
J_{2}(kr)\,  \hat{\sigma}(\xp,r)
\ee
with $K_{2}$ and $J_{2}$ are the Bessel functions. 
In papers \cite{Wusthoff:1997fz,Golec-Biernat:1999qd}, formula (\ref{eq:10}) was computed with two gluons exchanged between the diffractive system and the proton.
Then, the two gluon exchange interaction was substituted by the
dipole cross section $\hat\sigma=\hat{\sigma}(\xp,r)$ for the $q\qbar$ dipole interaction with the proton. 
For example, for the GBW parameterisation of the dipole cross section \cite{Wusthoff:1997fz}, which we discuss in the next section, is given by
\be
\sigmahat\equiv\sigmahat_{q\qbar}=\sigma_0\left(1-\eto^{-r^2Q_s^2/4}\right)\,.
\ee

However, the $q\qbar g$ system was computed in the approximation when parton transverse momenta fulfil the condition $k_{Tq}\approx k_{T\qbar}\gg k_{Tg}$.
Thus, in the large $N_c$ approximation, it can be treated as a gluonic
color dipole $gg$.  Such a  dipole interacts with  the relative color factor $C_A/C_F$ with respect to  the $q\qbar$ dipole.
Therefore,  the two gluon exchange formula should be  eikonalized with this color factor absorbed into the exponent. For the GBW parameterisation, this  leads to the following gluon dipole cross section in eq.~(\ref{eq:11})
\be\label{eq:sgg}
\sigmahat\equiv\sigmahat_{gg}=\sigma_0\left(1-\eto^{-(C_A/C_F) r^2Q_s^2/4}\right)\,.
\ee
In such a case, the color factor $C_A/C_F=9/4$ (for $N_c=3$) disappears from the normalisation of the scattering amplitude and we have to rescale the structure function in the following way
\be\label{eq:sgga}
F_T^{(q\qbar g)}~\to~\frac{1}{(C_A/C_F)^2}\,F_T^{(q\qbar g)}\,.
\ee
By the comparison with HERA data, 
we will show in the next section that the latter possibility is more appropriate
for the data description.
In this analysis we  considered two parameterisations: GBW from \cite{Golec-Biernat:1998js} and CGC parameterisation with heavy quarks \cite{Soyez:2007kg,Marquet:2007nf} see our paper \cite{ GolecBiernat:2008gk} for more details. 


\section{Diffractive charm quark production}
\label{sec:charm}

\begin{wrapfigure}{r}{0.5\columnwidth}
\vskip -1.cm
\centerline{\includegraphics[width=0.55\columnwidth]{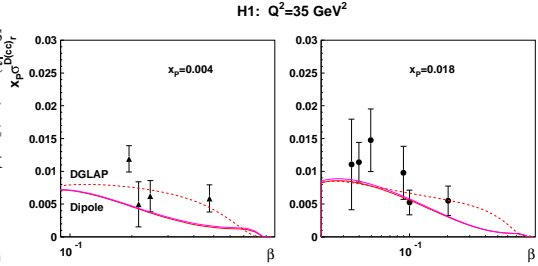}}
\caption{A comparison of the collinear factorisation predictions with the GBW and
CGC gluon distributions (solid lines) with the HERA data on the open diffractive charm production}
\label{Fig:1}
\end{wrapfigure}


In the diffractive scattering heavy quarks are produced 
in quark-antiquark pairs, $c\cbar$ and $b\bbar$ for charm and bottom, respectively. 
Such pairs can be produced provided that 
the diffractive mass of is above the quark pair production threshold
\be\label{eq:12}
M^2=Q^2\left(\frac{1}{\beta}-1\right) > 4m_{c,b}^2
\ee
In the lowest order the diffractive state consist only the  $c\cbar$ or $b\bbar$ pair.
In the forthcoming we consider only charm production since bottom production is negligible.The corresponding contributions to $F_2^D$ are given by eqs.~(\ref{eq:5a}) and 
(\ref{eq:5}) with one flavour component.
For example, for charm production from transverse photons we have 
\beeq\nonumber
\label{eq:5new}
\xp F_T^{(c\cbar)}\!\!\eq\!\!
\frac{3 Q^4 e_c^2}{64\/\pi^4\beta B_d} \int_{z_{c}}^{1/2}
dz\, z(1-z)
\\
&\times&\left\{
[z^2+(1-z)^2]\,Q^2_c\,\phi_1^2 \,+\,m_c^2\, \phi_0^2
\right\}
\eeeq
where $m_c$ and $e_c$ are charm quark mass and electric charge, respectively.
The minimal value of diffractive mass equals:
$M^2_{min}=4m_c^2$, thus the maximal value of $\beta$ is given by
\be
\beta_{max}=\frac{Q^2}{Q^2+4m_c^2}\,.
\ee
In such a case,  $z_c=1/2$ in Eq., for $\xp F_T^{(c\cbar)}$
and  $F_{T,L}^{(c\cbar)}=0$ for $\beta> \beta_{max}$. Figure \ref{Fig:1} shows the collinear factorisation predictions for the diffractive charm production confronted with the new HERA data \cite{Aktas:2006up} on the charm component of the reduced cross section.
The solid curves, which are barley distinguishable, correspond to the 
result with the GBW and CGC parameterisations of the diffractive gluon distributions.
The dashed lines are computed for the gluon distribution from a fit  to the H1 data \cite{GolecBiernat:2007kv} based on the DGLAP equations. 

\section{Comparison with the HERA data}
\label{sec:f2d}
\begin{wrapfigure}{r}{0.5\columnwidth}
\vskip -1.0cm
\centerline{\includegraphics[width=0.55\columnwidth]{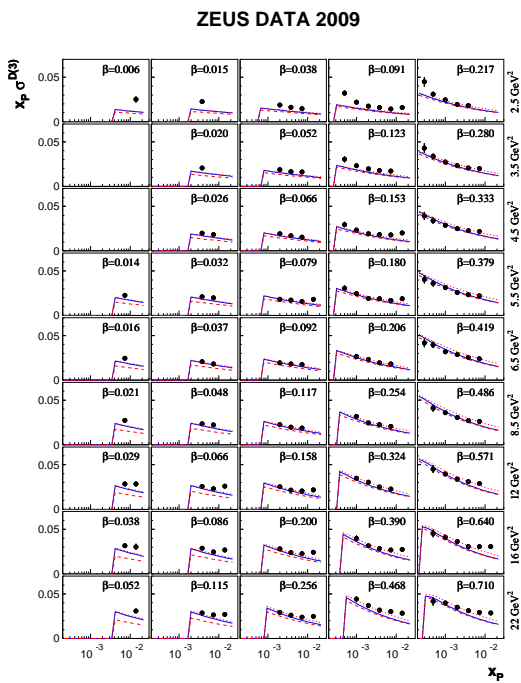}}
\caption{A comparison of $\sigma_r^D$ from the two considered dipole models with the newest ZEUS collaboration data \cite{Chekanov:2008fh}. The solid lines correspond to the GBW parameterisation of the dipole cross section  with the color factor modifications (\ref{eq:sgg}) and (\ref{eq:sgga}), while the dotted lines correspond to the CGC parameterisation.  The dashed lines show the results without the charm contribution.}
\label{fig:4ZEUSa}
\end{wrapfigure}

Figure \ref{fig:4ZEUSa} 
shows a comparison of the dipole
model predictions with the ZEUS collaboration data \cite{Chekanov:2008fh} on  the reduced cross section
\be
\sigma_r^{D}=F_2^{D}-\frac{y^2}{1+(1-y)^2}F_L^{D}\,.
\ee
We included  the charm contribution in the above structure functions.
The solid lines correspond to the GBW parameterisation of the dipole cross section with the color factor modifications (\ref{eq:sgg}) and (\ref{eq:sgga}) of the $q\qbar g$ component, while the dashed lines are obtained from the CGC parameterisation. We see that the two sets of curves are barely distinguishable
The color factor modification of the $q\qbar g$ component in the GBW parameterisation is necessary since the curves without such a modification significantly overshoot the data
(by a factor of two or so)
in the region of small $\beta$ where the ${q\qbar g}$ component dominates.
The comparison  of the predictions with the data also reveals a very important aspect of  the three component dipole model  (\ref{eq:1}). In the small $\beta$ region, the  curves are systematically  below the data points, which effect  may be attributed to the lack of higher order components in the diffractive state, 
i.e. with more than one gluon or $q\qbar$ pair.  This is also seen for the H1 collaboration data \cite{Aktas:2006hy} shown in our last paper \cite{GolecBiernat:2008gk}. In our analysis we also computed the gluon distributions respectively for the GBW parameterisation with the color factor modification and for the CGC  parameterisation. We compare them with the gluon distributions found in the DGLAP fit with higher twist to the recent H1 data \cite{GolecBiernat:2007kv} and  use it for the computation of the charm contribution to $F_2^D$, see our last work \cite{GolecBiernat:2008gk}.

\section{Summary}

We presented a comparison of the dipole model results with the GBW and CGC parameterisations on the diffractive structure functions with the HERA data. The three component model with the $q\qbar$ and $q\qbar g$ diffractive states  describe reasonable well  the recent data. 
However,  the region of small values of $\beta$ needs some refinement by considering components with more gluons and $q\qbar$ pairs in the diffractive state. This can be achieved in the DGLAP based approach which sums partonic emissions in the diffractive state in the transverse momentum ordering approximation. It is also important that the charm contribution, described in Sec.~\ref{sec:charm}, is added into the analysis. Without this contribution the comparison would be much worse than that shown here.



%


\end{document}